\documentclass[useAMS,usenatbib]{mn2e}

\usepackage{umlaut}
\usepackage{color}
\usepackage{graphicx}
\usepackage{graphics}
\usepackage{rotating}
\usepackage{amsmath}        
\usepackage{amssymb}        
\usepackage{epsfig}        

\title[Black hole formation in V404 Cyg]{The formation of the black
hole in the X-ray binary system V404 Cyg}

\author[J.C.A.~Miller-Jones et al.]
 {J.C.A.~Miller-Jones,$^{1,2}$\thanks{email: jmiller@nrao.edu}
 P.G.~Jonker,$^{3,4}$ G.~Nelemans,$^5$ S.~Portegies Zwart,$^{6,7}$ \and V.~Dhawan,$^8$ W.~Brisken,$^8$ E.~Gallo,$^{9,10}$ M.P.~Rupen$^8$\\
$^1$NRAO Headquarters, 520 Edgemont Road, Charlottesville,
 VA, 22903, USA\\
$^2$Jansky Fellow, National Radio Astronomy Observatory\\
$^3$SRON, Netherlands Institute for Space Research, 3584 CA Utrecht,
 the Netherlands\\
$^4$Harvard-Smithsonian Center for Astrophysics, Cambridge, MA 02138,
 USA\\
$^5$Department of Astrophysics, IMAPP, Radboud University, Toernooiveld 1, 6525 ED, Nijmegen, the Netherlands\\
$^6$Astronomical Institute `Anton Pannekoek', University of Amsterdam, Kruislaan 403, 1098 SJ Amsterdam, the Netherlands\\
$^7$Section Computational Science, University of Amsterdam, Kruislaan 403, 1098 SJ Amsterdam, the Netherlands\\
$^8$NRAO, Array Operations Center, 1003 Lopezville Road, Socorro, NM
 87801, U.S.A.\\
$^9$Physics Department, Broida Hall, University of California, Santa
 Barbara, CA, 93106, USA\\
$^{10}$Chandra Fellow\\
}
\begin{document}

\date{Accepted 2008 December 15.  Received 2008 December 11; in original form 2008 November 3}

\pagerange{\pageref{firstpage}--\pageref{lastpage}} \pubyear{2008}

\maketitle

\label{firstpage}

\begin{abstract}
Using new and archival radio data, we have measured the proper motion
of the black hole X-ray binary V404 Cyg to be
$9.2\pm0.3$\,mas\,yr$^{-1}$.  Combined with the systemic radial
velocity from the literature, we derive the full three-dimensional
heliocentric space velocity of the system, which we use to calculate a
peculiar velocity in the range 47--102\,km\,s$^{-1}$, with a best
fitting value of 64\,km\,s$^{-1}$.  We consider possible explanations
for the observed peculiar velocity, and find that the black hole
cannot have formed via direct collapse.  A natal supernova is
required, in which either significant mass ($\sim 11M_{\odot}$) was
lost, giving rise to a symmetric Blaauw kick of up to
$\sim65$\,km\,s$^{-1}$, or, more probably, asymmetries in the
supernova led to an additional kick out of the orbital plane of the
binary system.  In the case of a purely symmetric kick, the black hole
must have been formed with a mass $\sim 9M_{\odot}$, since when it has
accreted 0.5--1.5\,$M_{\odot}$ from its companion.
\end{abstract}


\section{Introduction}
\label{sec:intro}

The proper motions of X-ray binary systems can be used to derive
important information on the birthplaces and formation mechanisms of
their compact objects.  Since typical proper motions are of order a
few milliarcseconds per year, high-resolution observations and long
time baselines are required to measure the transverse motions of such
systems across the sky.  Proper motions have only been measured for a
handful of X-ray binary systems to date
\citep{Mir01,Mir02,Rib02,Mir03,Mir03b,Dha06,Dha07}, and only one X-ray
binary, Sco X-1, has a measured proper motion, parallactic distance,
and radial velocity \citep{Bra99,Cow75}.  With the position, proper
motion, radial velocity, and distance to the system, the
full three-dimensional space velocity of the system can be derived.
Along with system parameters such as the component masses, orbital
period, donor temperature and luminosity, these parameters may be used
to reconstruct the full evolutionary history of the binary system back
to the time of compact object formation, as done for the systems GRO
J\,1655-40 \citep{Wil05} and XTE J\,1118+480 \citep{Fra08}.

By studying the distribution of black hole X-ray binary velocities
with compact object masses, we can derive constraints on theoretical
models of black hole formation \citep[e.g.][]{Fry01}.  The two most
common theoretical scenarios for creating a black hole involve either
a massive star collapsing directly into a black hole with very little
or no mass ejection, or delayed formation in a supernova, as fallback
onto the neutron star of material ejected during the explosion creates
the black hole.  In the latter case, constraints on the magnitude and
symmetry of any natal kick can also be derived.  When the primary star
reaches the end of its life and explodes as a supernova, the centre of
mass of the ejected material continues to move with the velocity the
progenitor had immediately before the explosion.  The centre of mass
of the binary system then recoils in the opposite direction.  Such a
kick \citep{Bla61} is thus constrained to lie in the orbital plane.
In addition, in the presence of asymmetries in the supernova
explosion, a further, asymmetric, kick, which need not be in the
orbital plane, may be imparted to the binary \citep[see, e.g.,][for
more detailed overviews]{Bra95a,Por98,Lai01}.  While pulsar proper
motions \citep{Lyn94} and the high eccentricities of Be/X-ray binaries
provide strong evidence for asymmetric kicks in neutron star formation
\citep{van97,van00}, less attention has been paid to black hole binary
systems.  In two of the black hole X-ray binaries with some of the
best observational constraints on the system parameters, XTE
J\,1118+480 and GRO J1655-40, there is evidence for an asymmetric
supernova kick \citep{Gua05,Wil05}.  There is also some evidence
suggesting an asymmetric kick in GRS\,1915+105 \citep{Dha07}, although
in that case the uncertainty in the distance to the system precludes
the derivation of strong constraints on the kick velocity.  However,
the black hole in Cygnus X-1 is inferred to have formed via direct
collapse \citep{Mir03}.  More black hole sources need to be studied to
measure black hole kicks and investigate consequent formation
mechanisms in order to constrain theoretical models of black hole
formation.

\subsection{V404 Cyg}
V404 Cyg is a dynamically-confirmed black hole X-ray binary system,
with a mass function of $6.08\pm0.06M_{\odot}$ \citep{Cas94}.  The
system comprises a black hole accretor of mass $12^{+3}_{-2}M_{\odot}$
\citep{Sha94} in a 6.5-d orbit with a $0.7^{+0.3}_{-0.2}M_{\odot}$ K0
IV stripped giant star \citep{Cas93,Kin93}.  The orbit is highly
circular, with an eccentricity of $e<3\times10^{-4}$, in agreement
with the fact that tidal forces just before the onset of mass transfer
should have recircularised the orbit following the supernova.  Its
low radial velocity \citep[$-0.4\pm2.2$\,km\,s$^{-1}$;][]{Cas94} has
been taken as evidence for a small natal kick \citep{Bra95,Nel99}.

In this paper we present measurements of the proper motion of V404
Cyg, and go on to derive its full three-dimensional space velocity,
infer its Galactocentric orbit, and discuss the implications for the
formation of the black hole in the light of the inferred natal kick
from the supernova.

\section{Observations and data reduction}
\label{sec:obs}
In order to investigate the proper motion of the source, we
interrogated the Very Large Array (VLA) archives for high-resolution
observations of V404 Cyg.  We selected only A-configuration
observations at frequencies of 8.4\,GHz and higher, to obtain the
highest possible astrometric accuracy.  We further restricted the
dataset to observations where the phase calibrator was J\,2025+3343,
the phase reference source used in the High Sensitivity Array (HSA)
observations of \citet{Mil08}.  Since all positions are measured
relative to the phase reference source, observations using a different
secondary calibrator would potentially have been subject to a
systematic positional offset.

The VLA data were reduced using standard procedures within the 31Dec08
version of {\sc aips} \citep{Gre03}.  A script was written in {\sc
ParselTongue}, the Python interface to {\sc aips}, to automate the
bulk of the calibration.  We corrected the derived positions for
shifts in the assumed calibrator position, using a reference position
of 20$^{\rm h}$25$^{\rm m}$10\fs8421050 33\degr43\arcmin00\farcs214430
(J\,2000) for the calibrator source.  All co-ordinates were precessed
to J\,2000 values using the {\sc aips} task {\sc uvfix}.  Using data
from the USNO ({\texttt
http://maia.usno.navy.mil/ser7/finals2000A.all}), we corrected for
offsets in (UT1-UTC), the difference between Universal Time (UT1) as
set by the rotation of the Earth and measured by very long baseline
interferometry (VLBI), and co-ordinated Universal Time (UTC), the
atomic time (TAI) adjusted with leap seconds.  Where measured offsets
were available, we also corrected for shifts in antenna positions
using the {\sc aips} task {\sc vlant}.  Source positions were measured
by fitting an elliptical Gaussian to the source in the image plane,
using the deconvolved, phase-referenced image prior to any
self-calibration.  The source was not resolved in any of the images.
We added an extra positional uncertainty of 10\,mas to the measured
VLA positions, to account for systematic uncertainties in the
astrometry.  The list of observations and derived source positions is
given in Table~\ref{tab:vla_obs}.

The dataset was enhanced by the use of two high-resolution VLBI
measurements of the source position.  We used the 8.4-GHz position of
\citet{Mil08}, and also obtained a second measurement using global
VLBI at 22\,GHz, under proposal code GM064.  Eight European stations
(Cambridge, Effelsberg, Jodrell Bank Mk{\sc ii}, Medicina, Metsahovi,
Noto, Onsala, and Robledo), all ten Very Long Baseline Array (VLBA)
stations, the phased VLA and the Green Bank Telescope (GBT)
participated in the experiment.  Data were taken from 21:30:00 {\sc
ut} on 2008 May 31 until 16:00:00 {\sc ut} on 2008 June 1, with the
European stations being on source for the first 12\,h of the run
(Robledo from 02:20:00 until 08:45:00 on 2008 June 1) and the North
American stations for the second 12\,h (Mauna Kea from 07:30:00, when
the source rose in Hawaii).  The overlap time, when both sets of
stations were on source, was 5.5\,h.  The phase reference and fringe
finder source was J\,2025+3343.  We observed with a total bit rate of
512\,Mb\,s$^{-1}$, with a bandwidth of 64\,MHz per polarization.  We
observed in 150-s cycles, spending 1.5\,min on the target and 1\,min
on the calibrator in each cycle.  The VLA was phased up at the start
of each calibrator scan, and we made referenced pointing observations
with the larger dishes (the VLA, GBT, Effelsberg and Robledo) every
1--2\,h.  Data were reduced using standard procedures within {\sc aips}.  We
detected the source at a significance level of $4.7\sigma$.

\begin{table*}
\begin{center}
\scriptsize
\begin{tabular}{cccccccccccc}
\hline\hline
Code & Date & MJD & Error & Config. & Frequency & RA & Error & Dec & Error\\
& & & (d) & & (GHz) & & (sec) & & (arcsec)\\
\hline
AH348 & 1990 Feb 16 & 47938.71 & 0.18 & A & 14.94 & 20$^{\rm h}$24$^{\rm m}$03\fs82854 & 0.00085 & 33\degr52\arcmin02\farcs0348 & 0.0103\\
AH385 & 1990 Mar 08 & 47958.70 & 0.12 & A & 14.94 & 20$^{\rm h}$24$^{\rm m}$03\fs82896 & 0.00083 & 33\degr52\arcmin02\farcs0379 & 0.0102\\
AH390 & 1990 Mar 25 & 47975.55 & 0.09 & A & 8.44 & 20$^{\rm h}$24$^{\rm m}$03\fs82865 & 0.00081 & 33\degr52\arcmin02\farcs0439 & 0.0101\\
AH424 & 1991 Sep 25 & 48525.03 & 0.05 & AB & 14.94 & 20$^{\rm h}$24$^{\rm m}$03\fs82733 & 0.00276 & 33\degr52\arcmin01\farcs9563 & 0.0235\\
AH424 & 1991 Sep 25 & 48525.04 & 0.05 & AB & 8.44 & 20$^{\rm h}$24$^{\rm m}$03\fs82321 & 0.00148 & 33\degr52\arcmin02\farcs0143 & 0.0136\\
AH424 & 1992 Oct 20 & 48916.03 & 0.07 & A & 14.94 & 20$^{\rm h}$24$^{\rm m}$03\fs82752 & 0.00096 & 33\degr52\arcmin02\farcs0208 & 0.0122\\
AH424 & 1992 Oct 20 & 48916.04 & 0.07 & A & 8.44 & 20$^{\rm h}$24$^{\rm m}$03\fs82673 & 0.00084 & 33\degr52\arcmin02\farcs0196 & 0.0105\\
AH390 & 1993 Jan 28 & 49015.68 & 0.04 & AB & 14.94 & 20$^{\rm h}$24$^{\rm m}$03\fs82779 & 0.00222 & 33\degr52\arcmin01\farcs9964 & 0.0173\\
AH390 & 1993 Jan 28 & 49015.68 & 0.02 & AB & 8.44 & 20$^{\rm h}$24$^{\rm m}$03\fs82427 & 0.00219 & 33\degr52\arcmin02\farcs0317 & 0.0158\\
AH641 & 1998 May 04 & 50937.60 & 0.04 & A & 8.46 & 20$^{\rm h}$24$^{\rm m}$03\fs82493 & 0.00086 & 33\degr52\arcmin01\farcs9820 & 0.0109\\
AH669 & 1999 Jul 04 & 51363.41 & 0.01 & A & 8.46 & 20$^{\rm h}$24$^{\rm m}$03\fs82479 & 0.00144 & 33\degr52\arcmin01\farcs9721 & 0.0156\\
AH669 & 1999 Jul 13 & 51372.37 & 0.01 & A & 8.46 & 20$^{\rm h}$24$^{\rm m}$03\fs82431 & 0.00115 & 33\degr52\arcmin01\farcs9984 & 0.0143\\
AH669 & 1999 Jul 26 & 51385.30 & 0.01 & A & 8.46 & 20$^{\rm h}$24$^{\rm m}$03\fs82508 & 0.00260 & 33\degr52\arcmin01\farcs9470 & 0.0343\\
AH669 & 1999 Aug 22 & 51412.18 & 0.01 & A & 8.46 & 20$^{\rm h}$24$^{\rm m}$03\fs82443 & 0.00163 & 33\degr52\arcmin01\farcs9429 & 0.0176\\
AH669 & 1999 Sep 01 & 51422.21 & 0.01 & A & 8.46 & 20$^{\rm h}$24$^{\rm m}$03\fs82417 & 0.00152 & 33\degr52\arcmin02\farcs0063 & 0.0207\\
AH669 & 1999 Sep 04 & 51426.02 & 0.01 & A & 8.46 & 20$^{\rm h}$24$^{\rm m}$03\fs82397 & 0.00194 & 33\degr52\arcmin01\farcs9464 & 0.0188\\
AH669 & 2000 Oct 20 & 51837.33 & 0.01 & A & 8.46 & 20$^{\rm h}$24$^{\rm m}$03\fs82718 & 0.00595 & 33\degr52\arcmin01\farcs9676 & 0.0405\\
AR476 & 2002 Feb 03 & 52308.69 & 0.01 & A & 8.46 & 20$^{\rm h}$24$^{\rm m}$03\fs82325 & 0.00088 & 33\degr52\arcmin01\farcs9481 & 0.0108\\
AR476 & 2002 Mar 01 & 52334.54 & 0.01 & A & 8.46 & 20$^{\rm h}$24$^{\rm m}$03\fs82586 & 0.00257 & 33\degr52\arcmin01\farcs9457 & 0.0224\\
AH823 & 2003 Jul 29 & 52849.30 & 0.29 & A & 8.46 & 20$^{\rm h}$24$^{\rm m}$03\fs82440 & 0.00101 & 33\degr52\arcmin01\farcs9317 & 0.0120\\
BG168 & 2007 Dec 02 & 54436.84 & 0.09 & HSA & 8.42 & 20$^{\rm h}$24$^{\rm m}$03\fs82129 & 0.000010 & 33\degr52\arcmin01\farcs8993 & 0.0003\\
GM064 & 2008 May 31 & 54618.42 & 0.25 & Global & 22.22 & 20$^{\rm h}$24$^{\rm m}$03\fs821082 & 0.000009 & 33\degr52\arcmin01\farcs8957 & 0.0001\\
\hline
\end{tabular}
\end{center}
{\caption{\label{tab:vla_obs}Summary of the VLA and VLBI observations.
    The errors on the MJD values are taken as half the length of the
    observation.  The quoted positional errors for the VLA
    observations are the sum in quadrature of the statistical errors
    and a systematic uncertainty of 10\,mas.  All co-ordinates are
    for epoch J2000.  The coordinate system is based on the assumed
    position of the calibrator source J\,2025+3343, taken to be
    (J\,2000) 20$^{\rm h}$25$^{\rm m}$10\fs8421050(224)
    33\degr43$^{\prime}$00\farcs21443(42).  The HSA and global VLBI
    observations have been corrected for parallax, assuming a source
    distance of 4\,kpc, and accounting for the uncertainty in the
    parallax when calculating the positional error bars.}}
\end{table*}

\section{Results}
\begin{figure}
\begin{center}
\includegraphics[width=\columnwidth]{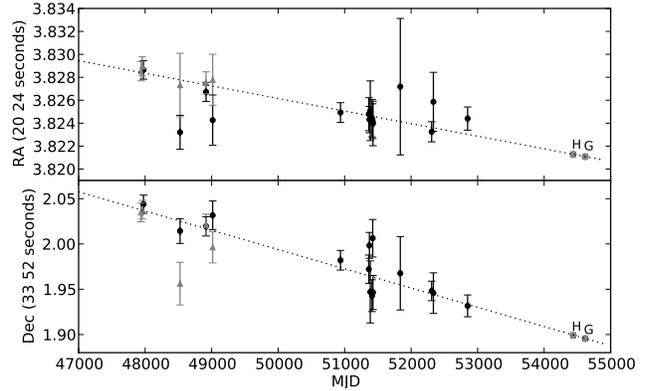}
\caption{Top panel: Measured Right Ascension as a function of time,
  from 1990 to 2007.  Bottom panel: Declination as a function of time.
  The dotted lines are the best fitting proper motions, $\mu_{\alpha}
  = (-4.99\pm0.19)\times10^{-4}$\,sec\,y$^{-1}$ in R.A. and
  $\mu_{\delta} = (-7.76\pm0.21)$\,mas\,y$^{-1}$ in Dec..  The black
  dots are the VLA 8.4\,GHz points and the light grey triangles are
  the 15-GHz VLA measurements.  The dark grey squares labelled `H' and
  `G' are from the 8.4-GHz HSA observations of \citet{Mil08} and the
  22-GHz global VLBI observations reported in this paper,
  respectively.
\label{fig:ra_dec}}
\end{center}
\end{figure}

Fig.~\ref{fig:ra_dec} shows the measured Right Ascension and
Declination as a function of time, over the $\sim20$ years since the
1989 outburst of V404 Cyg.  The best fitting proper motions, in Right
Ascension and Declination respectively, are
\begin{align}
\mu_{\alpha}\cos\delta &= -4.99\pm0.19\quad {\rm mas}\quad {\rm y}^{-1}\\
\mu_{\delta} &= -7.76\pm0.21\quad {\rm mas}\quad {\rm y}^{-1}.
\end{align}
Thus the total proper motion is $\mu = 9.2\pm0.3$\,mas\,y$^{-1}$.
All uncertainties are 68 per cent confidence limits, and unless
otherwise noted we will henceforth quote $1\sigma$ error bars on all
measurements.

The fit gives a reference position (prior to correcting for the effects
of the unknown parallax) of $20^{\rm h}24^{\rm m}03\fs82177(2)$
$33\degr52^{\prime}01\farcs9088(3)$ (J\,2000) on MJD\,54000.0, from
which we can use the proper motion to determine the predicted
position at any future time.

\subsection{Converting to Galactic Space Velocity co-ordinates}
With the systemic radial velocity of $-0.4\pm2.2$\,km\,s$^{-1}$
measured from optical H$\alpha$ studies \citep{Cas94}, we can
calculate, for a given source distance, the full three-dimensional
space velocity of the system.  The best constraint on the source
distance is $4.0^{+2.0}_{-1.2}$\,kpc \citep{Jon04}.  Using the
transformations of \citet{Joh87}, and the standard solar motion of
($U_{\odot}=10.0\pm0.36$, $V_{\odot}=5.25\pm0.62$,
$W_{\odot}=7.17\pm0.38$) km\,s$^{-1}$ \citep{Deh98}, we can compute
the heliocentric Galactic space velocity components $U$, $V$ and $W$
(defined as $U$ positive towards the Galactic Centre, $V$ positive
towards $l=90$\degr, and $W$ positive towards the North Galactic
Pole).  The derived system parameters are given in Table
\ref{tab:uvw}.
\begin{table}
\begin{center}
\begin{tabular}{ll}
\hline\hline
Parameter & Value\\
\hline
Galactic longitude $l$ & 73.12\degr\\
Galactic latitude $b$ & -2.09\degr\\
Distance $d$ (kpc) & $4.0^{+2.0}_{-1.2}$\\
Systemic velocity $\gamma$ (km\,s$^{-1}$) & $-0.4\pm2.2$\\
Proper motion $\mu_{\alpha}\cos\delta$ (mas\,y$^{-1}$) & $-4.99\pm0.19$\\
Proper motion $\mu_{\delta}$ (mas\,y$^{-1}$) & $-7.76\pm0.21$\\
$U$ (km\,s$^{-1}$) & $177.1\pm3.8^{+83.7}_{-50.2}$\\
$V$ (km\,s$^{-1}$) & $-46.1\pm2.4_{-25.5}^{+15.3}$\\
$W$ (km\,s$^{-1}$) & $0.2\pm3.7^{+2.1}_{-3.5}$\\
$U-<U>$ (km\,s$^{-1}$) & $62.0^{+39.4}_{-17.4}$\\
$V-<V>$ (km\,s$^{-1}$) & $-16.1_{-0.0}^{+6.5}$\\
$W-<W>$ (km\,s$^{-1}$) & $0.2^{+2.1}_{-3.5}$\\
$v_{\rm pec}$ (km\,s$^{-1}$)& $64.1\pm3.7^{+37.8}_{-16.6}$\\
\hline
\end{tabular}
\end{center}
{\caption{\label{tab:uvw}Measured and derived parameters.  $U$, $V$
and $W$ are the Galactic space velocity components in the direction of
the Galactic Centre, $l=90^{\circ}$ and $b=90^{\circ}$ respectively.
The first set of error bars accounts for uncertainties in the measured
space velocities only, and the second takes into account the distance
uncertainty. $U-<U>$, $V-<V>$ and $W-<W>$ are the discrepancies from
the velocities that would be expected for Galactic rotation.  Summing
these discrepancies in quadrature gives the peculiar velocity, $v_{\rm
pec}$.  The major source of error in these values is the distance
uncertainty.}}
\end{table}

\begin{figure}
\includegraphics[width=\columnwidth]{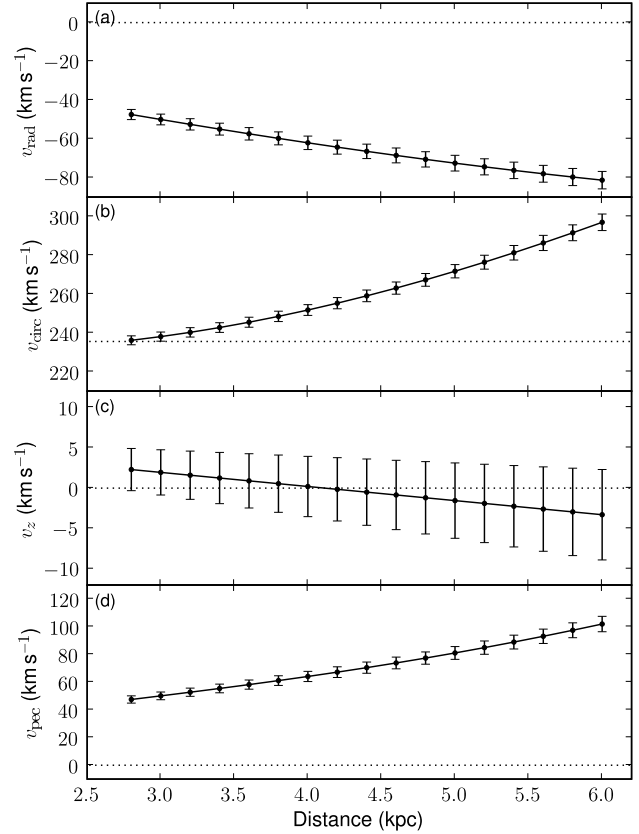}
\caption{Derived Galactocentric velocities of V404 Cyg, as a function
  of source distance.  (a) shows the radial velocity, $v_{\rm rad}$,
  (b) shows the circular velocity, $v_{\rm circ}$, (c) shows the
  velocity out of the Galactic plane, $W$, and (d) shows the peculiar
  velocity, $v_{\rm pec}$, i.e.\ the difference from the values
  expected for a source participating in the Galactic rotation.
  Dotted lines show the expected values of 0\,km\,s$^{-1}$ for the
  radial, $W$ and peculiar velocities, and 236\,km\,s$^{-1}$ for the
  circular velocity.  Error bars only account for uncertainties in the
  space velocity components, assuming zero error on the distance at
  each plotted point.\label{fig:v_pec}}
\end{figure}

For a given distance, the expected values of $U$ and $V$ can be
calculated, assuming the source participates in the Galactic rotation.
\citet{Rei04} determined the angular rotation rate of the LSR at the
Sun, $\Theta_0/R_0 = 29.45\pm0.15$\,km\,s$^{-1}$\,kpc$^{-1}$.
Assuming a Galactocentric distance of 8.0\,kpc \citep{Rei93}, this
implies a circular velocity of 236\,km\,s$^{-1}$.  For circular
rotation about the Galactic Centre, the $W$ component of the velocity
is expected to be zero.  As done by \citet{Dha07} for GRS\,1915+105,
we can transform the measured values of $U$ and $V$ into radial and
circular velocities about the Galactic Centre, expected to be $v_{\rm
rad} = 0$\,km\,s$^{-1}$ and $v_{\rm circ} = 236$\,km\,s$^{-1}$
respectively.  Fig.~\ref{fig:v_pec} shows the derived radial, circular
and $W$ velocities and the peculiar velocity of V404 Cyg.  The
peculiar velocity is defined to be the difference between the measured
3-dimensional space velocity and that expected for a source
participating in the Galactic rotation,
\begin{equation}
v_{\rm pec} = \left(v_{\rm rad}^2+(v_{\rm circ}-236)^2+W^2\right)^{1/2}
\label{eq:vpec}
\end{equation}

For the range of distances found by \citet{Jon04}, 2.8--6.0\,kpc, it
is clear that the peculiar velocity is non-zero.  We derive a value of
$v_{\rm pec} = 64.1\pm3.7^{+37.8}_{-16.6}$\,km\,s$^{-1}$, where the
first error bar accounts for statistical error in the space
velocities, and the second for the distance uncertainty.  The
predominant component of the peculiar velocity is radial, with a
circular velocity slightly faster than expected, and a velocity out of
the Galactic plane consistent with zero.

\section{The orbital trajectory in the Galactic potential}
\label{sec:orbit}
\begin{figure*}
\begin{center}
\includegraphics[width=\textwidth]{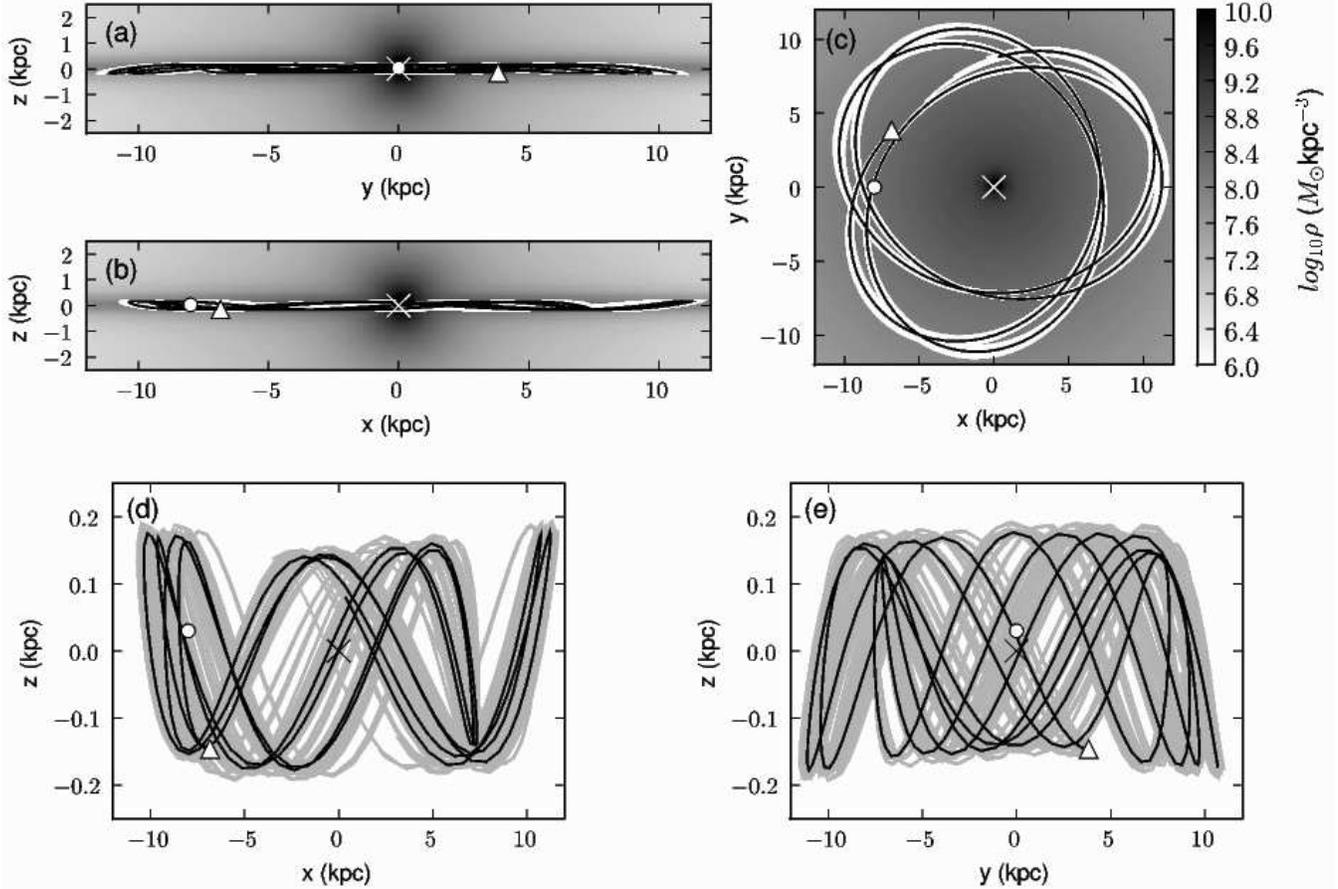}
\caption{The computed orbit of V404 Cyg over the last 1\,Gyr, using
  the Galactic potential of \citet{Joh95} and assuming a current
  source distance of 4.0\,kpc.  In (a), (b) and (c), the greyscale is
  a logarithmic representation of the Galactic mass density
  ($M_{\odot}$\,kpc$^{-3}$), the black line denotes the computed orbit
  for the best fitting space velocities and positions, and the white
  trace indicates the error due to uncertainty in the space velocity
  components.  For clarity, the effect of the distance uncertainty is
  not shown (see Fig.~\ref{fig:distances}).  Panels (d) and (e) are a
  zoomed-in version of the $x$--$z$ and $y$--$z$ planes respectively
  ($x$, $y$ and $z$ defined as the directions $l=0^{\circ}$,
  $l=90^{\circ}$, $b=90^{\circ}$ respectively), with the black line
  indicating the best-fitting trajectory and the grey trace once again
  indicating the spread due to uncertainty in the measured space
  velocities only.  In all panels, the cross marks the Galactic
  Centre, and the open circle and triangle mark the current positions
  of the Sun and V404 Cyg respectively.
\label{fig:gal_orbit}}
\end{center}
\end{figure*}

\begin{figure}
\begin{center}
\includegraphics[width=\columnwidth]{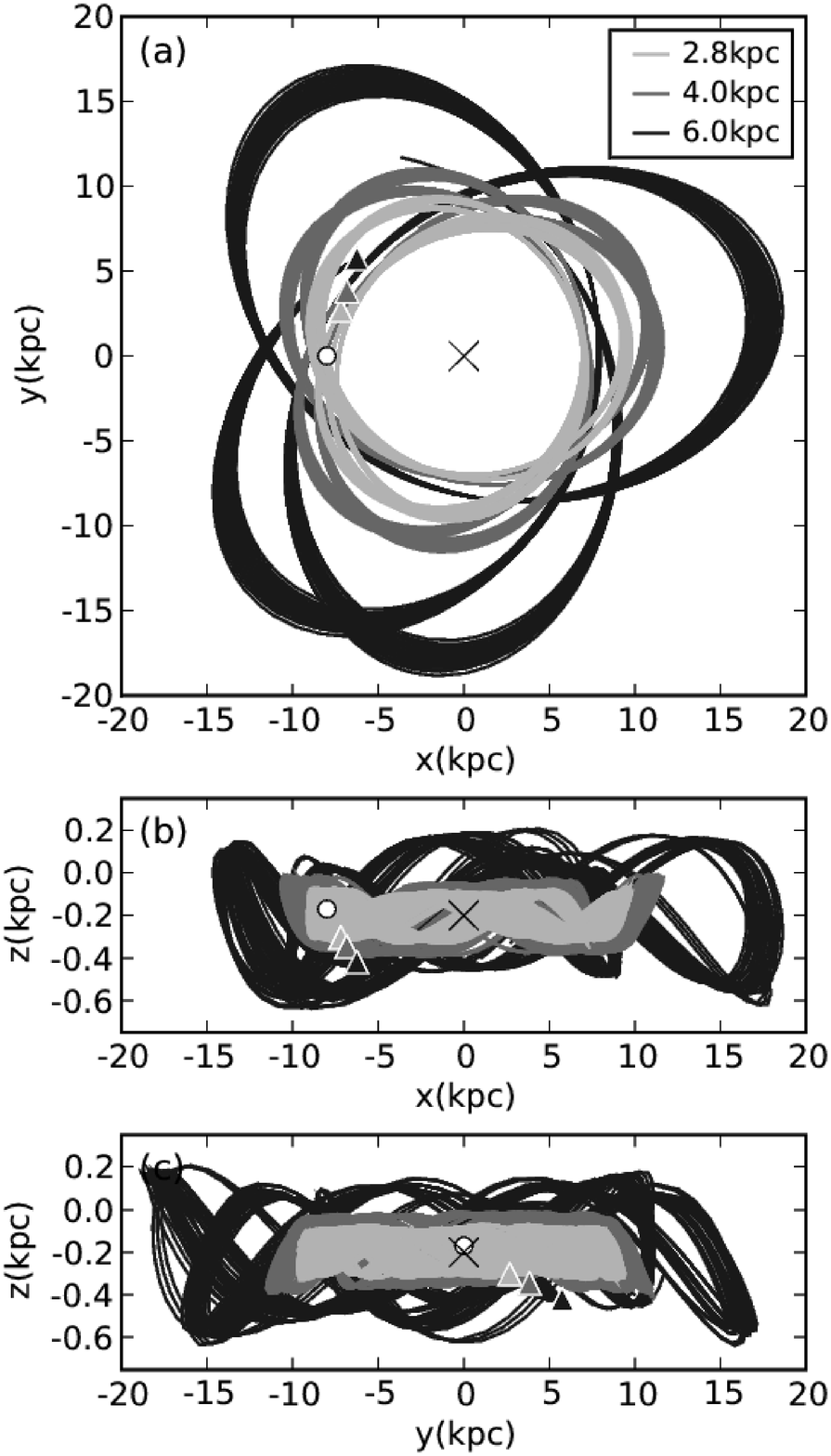}
\caption{The effect of the distance uncertainty on the computed
  Galactocentric orbit of V404 Cyg.  Trajectories have been computed
  for the minimum, best fitting and maximum distances derived by
  \citet{Jon04}, using the measured space velocities and the Galactic
  potential of \citet{Joh95}.  For each distance, the spread in the
  trajectories due to uncertainties in the space velocity components
  has been plotted.  In all panels, the cross marks the Galactic
  Centre, and the open circle and triangles mark the current positions
  of the Sun and V404 Cyg (at its different distances) respectively.
\label{fig:distances}}
\end{center}
\end{figure}

From the known source position and the measured spatial velocity
components, we can integrate backwards in time to compute the orbital
trajectory of the system in the potential of the Galaxy.  Using a
fifth-order Runge-Kutta algorithm \citep{Pre92} to perform the
integration, we compared the predictions of several different models
for the Galactic potential \citep[Carlberg \& Innanen 1987, using the
revised parameters of ][]{Kui89,Pac90,Joh95,Wol95,Fly96,deO02}, all
using some combination of one or more disc, spherical bulge and halo
components.  A representative orbit reconstruction using the model of
\citet{Joh95} is shown in Fig.~\ref{fig:gal_orbit}.
While the errors in the space velocity components make little
difference to the computed orbital trajectory for a given model, the
uncertainty in the distance has much more of an effect.
Fig.~\ref{fig:distances} compares the trajectories computed for 2.8,
4.0 and 6.0\,kpc, the lower, mean and upper bounds to the possible
range of distances.  A larger source distance implies a more
elliptical orbit, which reaches further from the Galactic Centre at
apogalacticon.

Comparing the different models for the Galactic potential, we find
that the predicted orbital trajectories in the Galactic plane begin to
diverge significantly after only 25--30\,Myr.  In the perpendicular
direction, the predictions diverge even faster, within 2--3\,Myr.
Given the uncertainty in the source distance and Galactic potential,
it is clearly impossible to integrate back in time over 0.4--0.8\,Gyr
(Section \ref{sec:kicks}) to locate the birthplace of the system.
However, for a given source distance, we can average over the ensemble
of model predictions to derive some generic properties of the orbit.
The eccentricity $e$, semi-major axis $a$, and the distances of the
apsides, $r_{\rm max}$ and $r_{\rm min}$, for the motion in the
Galactic plane, the maximum height reached above or below the plane,
$z_{\rm max}$, and the minimum and maximum values of the peculiar
velocity, $v_{\rm pec,min}$ and $v_{\rm pec,max}$, are given in Table
\ref{tab:orbit_parms}.

We can compare these values with those of the thick and thin disk
populations.  \citet{Bin98} give the velocity dispersions in the
radial, azimuthal and vertical directions (relative to the Galactic
plane) for both stellar populations, based on the data of
\citet{Edv93}, for stars within 80\,pc of the Sun.  A set of Monte
Carlo simulations established that the typical planar eccentricities
of the thin and thick disk populations were $0.12\pm0.06$ and
$0.29\pm0.15$ respectively, while the typical maximum height reached
above the plane was $0.17\pm0.14$ and $0.56\pm0.61$\,kpc respectively.
Unless V404 Cyg is at the maximum possible distance, it is likely to
have originated as a member of the thin disk population (as also
suggested by the age, metallicity and component masses of the system),
and received some sort of kick which increased the planar eccentricity
and the component of velocity out of the Galactic plane.

\begin{table}
\begin{center}
\begin{tabular}{lccc}
\hline\hline
Distance & 2.8\,kpc & 4.0\,kpc & 6.0\,kpc\\
\hline
$r_{\rm max}$ (kpc) & $9.8\pm0.4$ & $11.7\pm0.7$ & $20.2\pm4.4$\\
$r_{\rm min}$ (kpc) & $7.0\pm0.1$ & $7.2\pm0.1$ & $7.9\pm0.1$\\
$z_{\rm max}$ (kpc) & $0.13\pm0.02$ & $0.20\pm0.03$ & $0.47\pm0.14$\\
$a$ (kpc) & $8.4\pm0.2$ & $9.4\pm0.4$ & $14.1\pm2.2$\\
$e$ & $0.16\pm0.02$ & $0.24\pm0.03$ & $0.42\pm0.07$\\
$v_{\rm pec,min}$ (km\,s$^{-1}$) & $21.9\pm5.1$ & $38.9\pm6.4$ & $82.4\pm7.4$\\
$v_{\rm pec,max}$ (km\,s$^{-1}$) & $56.5\pm4.9$ & $78.8\pm5.7$ & $134.3\pm10.3$\\
\hline
\end{tabular}
\end{center}
{\caption{\label{tab:orbit_parms}Derived parameters of the
    Galactocentric orbit, averaged over the ensemble of models for the
    Galactic potential.  The parameters are the maximum and minimum
    distances from the Galactic Centre measured in the plane ($r_{\rm
    max}$ and $r_{\rm min}$), the maximum distance reached above or
    below the plane ($z_{\rm max}$), the semi-major axis $a$, the
    orbital eccentricity $e$ (both calculated in the plane), and the
    minimum and maximum peculiar velocity ($v_{\rm pec,min}$ and
    $v_{\rm pec,max}$ respectively).  Uncertainties are the scatter
    due to both the errors on the measured velocities and the
    differing models for the Galactic Potential.}}
\end{table}

\section{The peculiar velocity}
\label{sec:kicks}
\begin{figure}
\begin{center}
\includegraphics[width=\columnwidth]{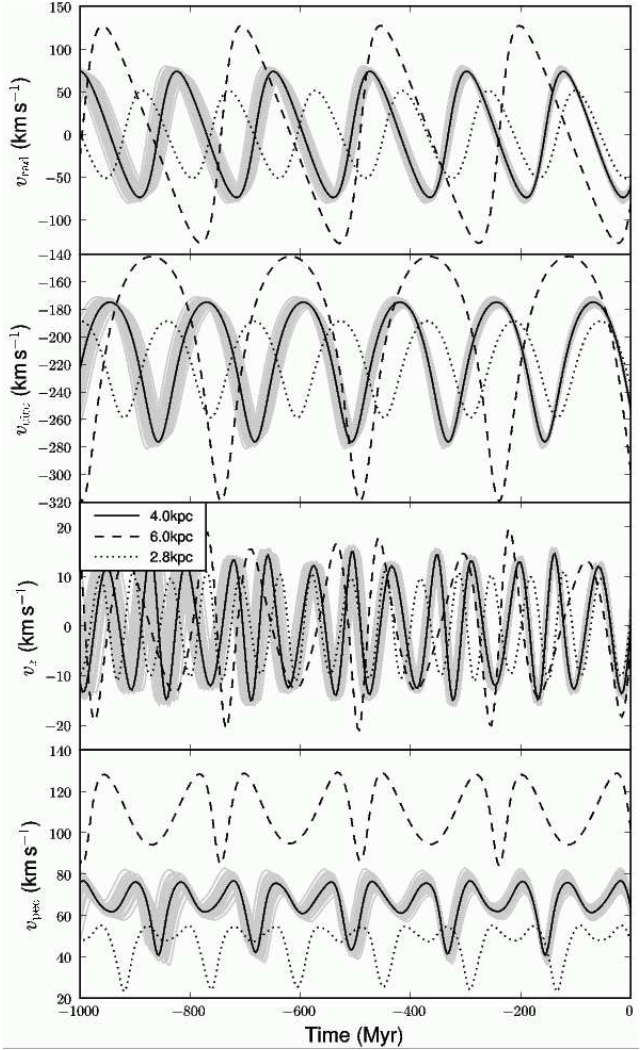}
\caption{Variation in Galactocentric radial velocity $v_{\rm rad}$,
  circular velocity $v_{\rm circ}$, perpendicular velocity $v_z$, and
  peculiar velocity $v_{\rm pec}$, as a function of time while
  integrating the orbit backwards over the last Gyr, using the
  potential of \citet{Joh95}.  Solid, dotted and dashed lines are for
  source distances of 4.0, 2.8 and 6.0\,kpc respectively.  The grey
  lines show the uncertainty arising from the error bars on the
  measured space velocities, for a distance of 4.0\,kpc.  The peculiar
  velocity varies as a function of time.
\label{fig:velocities}}
\end{center}
\end{figure}

The current peculiar velocity of the system is
$64.1\pm3.7^{+37.8}_{-16.6}$\,km\,s$^{-1}$.  However, owing to its
orbit in the Galactic potential, this is not a conserved quantity.
For a distance of 4\,kpc, we find that the peculiar velocity varies
between 39 and 79\,km\,s$^{-1}$ (see Table \ref{tab:orbit_parms} and
Fig.~\ref{fig:velocities}).  We go on to examine potential
explanations for this peculiar velocity.

\subsection{Symmetric supernova kick}

If it formed with a natal supernova, the system can receive a Blaauw
kick \citep{Bla61}, whereby the binary recoils to conserve momentum
after mass is instantaneously ejected from the primary.  For the
binary to remain bound after the supernova, the ejected mass $\Delta
M$ must be less than half the total mass of the system.  A maximum
ejected mass translates to a maximum recoil velocity of the binary,
for which an expression was derived by \citet{Nel99},
\begin{multline}
v_{\rm max} = 213 \left(\frac{\Delta M}{M_{\odot}}\right)
\left(\frac{m}{M_{\odot}}\right) \left({\frac{P_{\rm re-circ}}{{\rm
d}}}\right)^{-1/3}\\ 
\times\left(\frac{M_{\rm BH}+m}{M_{\odot}}\right)^{-5/3} {\rm km\,s}^{-1},
\label{eq:vmax}
\end{multline}
where $M_{\rm BH}$ and $m$ are the masses of the black hole and the
secondary immediately after the supernova, before mass transfer has
begun, and $P_{\rm re-circ}$ is the period of the orbit once it has
recircularized following the supernova, related to the orbital period
immediately after the supernova has occurred by $P_{\rm
re-circ}=P_{\rm post-SN}(1-e_{\rm post-SN}^2)^{3/2}$.
Re-circularization occurs before the start of mass transfer from the
secondary to the black hole, due to the strong tidal forces present
once the secondary has evolved and expanded sufficiently to come close
to filling its Roche Lobe.

\begin{figure}
\includegraphics[width=\columnwidth]{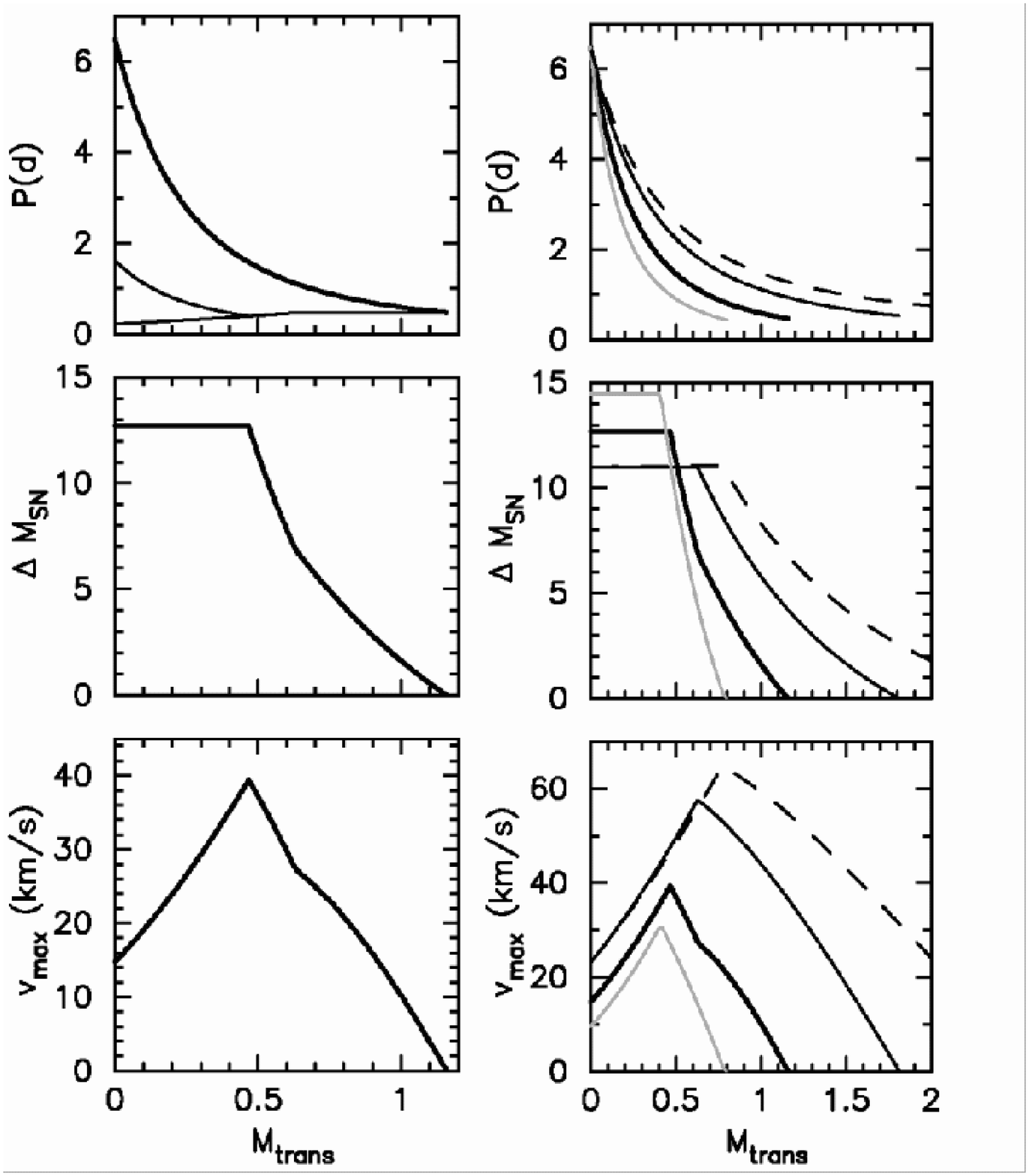}
\caption{Evolution of the post-supernova binary orbital period (top
  panels), maximum possible mass ejected in the supernova (middle
  panels), and maximum possible velocity kick (lower panels), as a
  function of total mass transferred since the supernova from the
  secondary to the black hole.  Left-hand panels are all for the case
  of conservative mass transfer, for current component masses of
  $(M_{\rm BH}, m) = $ (12, 0.7) $M_{\odot}$.  The top left plot shows
  the post-supernova orbital period (thick solid line), the
  pre-supernova orbital period (middle line) and the minimum permitted
  pre-supernova orbital period in which neither the helium star
  progenitor nor the main sequence companion fills its Roche lobe
  (bottom line).  The right-hand panels show the situation for
  conservative mass transfer using different current component masses,
  and one case of non-conservative mass transfer.  Thick solid lines
  are for current component masses of $(M_{\rm BH}, m) = $ (12, 0.7)
  $M_{\odot}$, thin solid lines are for (10, 1) $M_{\odot}$ and grey
  lines are for (14, 0.5) $M_{\odot}$, assuming conservative mass
  transfer in all cases.  Dashed lines indicate non-conservative mass
  transfer for the (10, 1) $M_{\odot}$ case, with ten per cent of the
  mass transferred being lost from the system.
\label{fig:vmax}}
\end{figure}

During its X-ray binary phase, the system undergoes mass transfer from
the donor star to the black hole, increasing the black hole mass and
orbital period and reducing the donor mass.  In the case of
conservative mass transfer, angular momentum conservation and Kepler's
Third Law imply that these parameters evolve as
\begin{equation}
P\propto(M_{\rm BH}m)^{-3}.
\label{eq:P-evolution}
\end{equation}
While we have no exact constraint on how long ago mass transfer began,
we can use the current orbital period and component masses
\citep{Sha94} together with Equation \ref{eq:P-evolution} to find the
system parameters at any point in the past, and the implied maximum
ejected mass and recoil velocity of a supernova which would have
created a system with those parameters (from Equation \ref{eq:vmax}).
This is shown in Fig.~\ref{fig:vmax} for three possible current
configurations of donor and accretor mass.

As mass is transferred from donor to accretor, the orbital period
increases (Equation \ref{eq:P-evolution}).  Thus for a larger
transferred mass, $M_{\rm trans}$, (a smaller initial black hole mass
and a less recent onset of mass transfer) the orbital period at the
onset of mass transfer is smaller.  The pre-supernova orbital period,
which is smaller due to the mass loss in the explosion (middle line in
the top left hand panel of Fig.~\ref{fig:vmax}), cannot be so short
that either the black hole progenitor or its companion fills its Roche
lobe, setting a minimum possible pre-supernova orbital period (lower
line in the top left hand panel of Fig.~\ref{fig:vmax}).  For small
values of $M_{\rm trans}$, the maximum mass lost in the supernova is
equal to half the total system mass, and the maximum kick velocity
$v_{\rm max}$ increases via Equation \ref{eq:vmax} because both the
post-supernova donor mass increases and the post-supernova period
decreases with increasing $M_{\rm trans}$.  Where the period
constraint becomes important, the maximum mass lost in the supernova
decreases, being set by the shortest allowed pre-supernova orbital
period.  The decreasing mass loss in the explosion then offsets the
increasing post-supernova donor mass, and the maximum possible kick
velocity decreases as $M_{\rm trans}$ increases further.

Equation \ref{eq:P-evolution} is valid only for the case of
conservative mass transfer.  However, there are strong indications
that this assumption is not valid for V404 Cyg.  At the very least,
the radio outbursts \citep[e.g.][]{Han92} suggest that material is
being lost to jet outflows.  \citet{Pod03} calculated binary evolution
tracks for a $10M_{\odot}$ black hole in orbit with donors of mass
2--$17M_{\odot}$ (for there to have been any symmetric kick,
Fig.~\ref{fig:vmax} shows that the transferred mass must be
$<2.5M_{\odot}$, implying that the initial secondary must have been
less massive than $\sim3.5M_{\odot}$), some of which they found could
reproduce the system parameters of V404 Cyg extremely well.  They took
into account non-conservative mass transfer, whereby mass transferred
in excess of the Eddington rate is lost from the system.  A comparison
with the equations for conservative mass transfer suggested that the
orbital period increases more slowly in the non-conservative case, by
a factor of at most 2.  As an example of how this could affect the
maximum Blaauw kick velocity, the dashed lines in Fig.~\ref{fig:vmax}
are for the current case of a $10M_{\odot}$ accretor with a
$1M_{\odot}$ donor, with 10 per cent of the transferred mass being
lost from the system and the orbital period increasing a factor 2 more
slowly than predicted by Equation \ref{eq:P-evolution}.

Our calculations show that for V404 Cyg, it is just possible to
achieve a kick of 64\,km\,s$^{-1}$ using only symmetric mass loss in
the supernova (a Blaauw kick).  If the explosion occurred at a point
in the Galactocentric orbit where the peculiar velocity was minimized
(Fig.~\ref{fig:velocities}), this could be sufficient to explain the
observed peculiar velocity, particularly if the system is towards the
lower end of the possible range of distances.  However, to achieve
such a large kick requires fine-tuning the parameters, to give a
relatively low initial black hole mass (closer to $9.2M_{\odot}$ than
the best fitting value of $12M_{\odot}$) and an ejection of
$\sim11M_{\odot}$ during the supernova.  Such a large mass loss, in
combination with the system parameters (a $\sim9M_{\odot}$ black hole
with a low-mass donor), do not make this scenario very plausible
\citep[see, e.g.,][for theoretical estimates of the amount of mass
ejected in a supernova as a function of progenitor mass; to form a
$9M_{\odot}$ black hole requires a progenitor of 25--30\,$M_{\odot}$, in
which case the mass ejected in the supernova is expected to be
$<3M_{\odot}$]{Fry01}.  We find it unlikely that a symmetric kick
alone is sufficient to explain the observed peculiar velocity.

\subsection{Velocity dispersion in the disc}
\label{sec:diffusion}
The donor star in V404 Cyg is a K0 subgiant \citep{Cas94} of mass
$0.7^{+0.3}_{-0.2}M_{\odot}$ \citep{Sha94}.  \citet{Kin93}
demonstrated that the system is a stripped giant, which evolves on the
nuclear timescale of the donor star, which is in the range
0.4--0.8\,Gyr.  Thus the secondary was initially significantly more
massive, and has transferred mass to the black hole.  From
Fig.~\ref{fig:vmax}, the donor star is unlikely to have transferred
more than $2.5M_{\odot}$ to the black hole during the mass transfer
phase, implying an initial donor mass of $<3.5M_{\odot}$.  The
evolutionary tracks of \citet{Pod03} show that such a system would
take of order 0.7--0.8\,Gyr to evolve to an orbital period of 6.5\,d,
in agreement with the range given by \citet{Kin93}.  In 0.8\,Gyr, the
system has made 3.5--5 orbits in the potential of the Galaxy
(depending on the model used for the Galactic potential; see Section
\ref{sec:orbit}), so could well have received some component of
peculiar velocity in the Galactic plane owing to non-axisymmetric
forces such as scattering from the potentials of spiral arms or
interstellar cloud complexes \citep{Wie77}.  However, estimates of the
velocity dispersion of the thin disk population \citep{Deh98,Mig00}
show that for F0--F5 stars (with initial masses comparable to that
estimated for the donor star in V404 Cyg prior to the onset of mass
transfer), it is of order 28\,km\,s$^{-1}$.

\citet{Mig00} fitted the Sun's peculiar motion and the differential
velocity field caused by the Galactic rotation to the parallaxes and
proper motions measured by Hipparcos.  Examining the fit residuals
showed the peculiar velocities to follow a three-dimensional Gaussian
distribution, with the predicted scatter in the velocity in Galactic
longitude being given by $(0.7<u^2>)^{1/2}$, where $<u^2>$ is the
velocity dispersion in the radial direction, determined as
$<u^2>^{1/2}=22.5\pm0.3$\,km\,s$^{-1}$ for F0-F5 stars.  The measured
peculiar velocity of V404 Cyg in the longitudinal direction is
$-64.0^{+18.5}_{-37.7}$\,km\,s$^{-1}$, implying a probability of
$7\times10^{-4}$ of being caused by Galactic velocity diffusion if the
source is at 4\,kpc, and only $1.6\times10^{-2}$ even if the source is
at 2.8\,kpc.  Thus this mechanism is unlikely to account for the
observed and inferred range of peculiar velocities.

\subsection{An asymmetric supernova kick?}
If neither the effects of stellar diffusion nor a symmetric kick can
explain the observed peculiar velocity, it could instead be the effect
of an asymmetric kick during the supernova.  The smaller dispersion in
$z$-distance (distance above or below the Galactic plane) of black
hole X-ray binaries when compared to neutron star systems had been
interpreted as evidence for smaller kicks when forming black holes
\citep{Whi96}.  However, \citet{Jon04}, using a larger sample and
revised distance estimates for the sources, found no evidence for such
a discrepancy, suggesting that black holes can receive natal kicks
comparable with those seen in neutron star systems.  Recent analysis
of the space velocities of the black hole X-ray binaries XTE
J\,1118+480 \citep{Gua05} and GRO J\,1655-40 \citep{Wil05} found
evidence for asymmetric kicks in these two systems, with such a kick
being mandated in the case of XTE J\,1118+480 \citep{Fra08}.

\begin{figure}
\begin{center}
\includegraphics[width=\columnwidth]{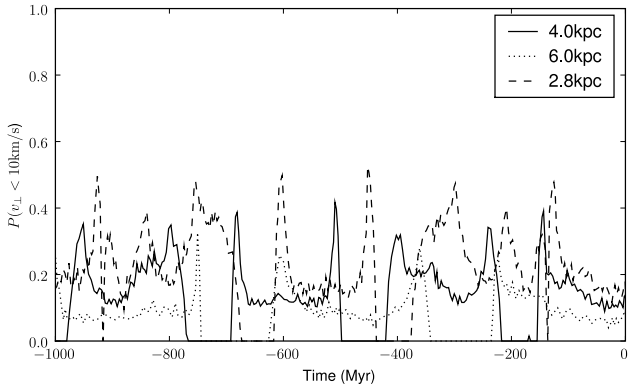}
\caption{Probability that the component of the peculiar velocity
  perpendicular to the orbital plane is less than 10\,km\,s$^{-1}$.
  We assumed a uniform probability for the longitude of the ascending
  node, for angles in the range $0<\Omega<2\pi$, and derived the
  probabilities for the best-fitting distance (4.0\,kpc; solid line),
  and the upper (6.0\,kpc; dotted line) and lower (2.8\,kpc; dashed
  line) limits to the distance.  A low probability implies that it is
  likely that the velocity perpendicular to the orbital plane is high,
  so the system has received an asymmetric kick at formation.
\label{fig:asymmetric_kick}}
\end{center}
\end{figure}

An asymmetric kick is not constrained to lie in the orbital plane, as
in the case of a Blaauw kick.  While the inclination angle of the
system to the line of sight is well-constrained to be
$56^{\circ}\pm4^{\circ}$ \citep{Sha94}, the longitude of the ascending
node, $\Omega$, with respect to the sky plane is not known, so we
cannot determine the absolute orientation of the binary orbital plane.
However, for a given value of $\Omega$, we can determine the
orientation of the orbital plane with respect to the Galactic axes
$x$, $y$ and $z$ (corresponding to the velocity components $U$, $V$
and $W$).  We assume that the orientation of the orbital plane does
not change with time.  For any point during its Galactocentric orbit,
we can use the positions and velocities computed in Section
\ref{sec:orbit} to calculate the component of the peculiar motion
perpendicular to the orbital plane, $v_{\perp}$, which provides a
lower limit to the asymmetric kick velocity should the supernova have
occurred at that point in time.  Assuming all values $0<\Omega<2\pi$
are equally likely, we can run Monte Carlo simulations to find the
probability that the component of the peculiar velocity perpendicular
to the orbital plane is equal to or less than would be expected for
the progenitor from the typical velocity dispersion of massive stars
\citep[taken as 10\,km\,s$^{-1}$ in one component; see][]{Mig00}.
Fig.~\ref{fig:asymmetric_kick} shows the probability that $v_{\perp}$
is less than 10\,km\,s$^{-1}$ for the mean and extreme values of the
distance, as a function of time.  The probability is low, of order
10--20 per cent, with little variation in the mean probability as a
function of source distance.  It is therefore unlikely (although
possible, except for certain short periods of time) that the measured
peculiar velocity perpendicular to the binary orbital plane can be
attributed to the velocity dispersion of the system prior to the
supernova.  Since a symmetric kick cannot give rise to velocity out of
the orbital plane, it is probable that there was an asymmetric kick
during the formation of the black hole.

\section{Discussion}
It appears that a supernova is required to explain the peculiar
velocity of V404 Cyg.  Thus the black hole in this system did not form
via direct collapse.  From the range of peculiar velocities inferred
during the Galactocentric orbit of the system, a symmetric supernova
kick could just be sufficient to explain the observations if the
source is at the lower end of the allowed range of distances.
However, the mass loss required, as well as the component of peculiar
velocity inferred to lie perpendicular to the orbital plane, make it
likely that the system was subject to an asymmetric kick during black
hole formation.

The full three-dimensional space velocities have been measured for a
handful of other black hole systems.  XTE J\,1118+480 \citep{Mir01}
and GRO J\,1655-40 \citep{Mir02} are both in highly eccentric orbits
around the Galactic Centre.  An asymmetric natal kick is required for
XTE J\,1118+480 \citep{Fra08}, and believed to have been likely in GRO
J\,1655-40 \citep{Wil05}.  The high-mass X-ray binary Cygnus X-1
\citep{Mir03} is moving at only $9\pm2$\,km\,s$^{-1}$ with respect to
its parent association Cyg OB3, implying that $<1M_{\odot}$ was
ejected in the natal supernova, and that the black hole formed by
direct collapse.  GRS\,1915+105 on the other hand, has been
  inferred to have an orbit and peculiar velocity \citep{Dha07}
fairly similar to that of V404 Cyg.  \citet{Dha07}, using a
  Galactocentric distance of $R_0=8.5$\,kpc and a circular velocity of
  $\Theta_0=220$\,km\,s$^{-1}$ for the LSR, found that a symmetric
supernova kick and stellar diffusion were insufficient to explain the
peculiar velocity unless the source was located at 9--10\,kpc, where
the peculiar velocity is minimized.  However, using the values of
  $R_0=8.0$\,kpc and $\Theta_0=236$\,km\,s$^{-1}$ assumed in this
  paper gives a minimum peculiar velocity of 23\,km\,s$^{-1}$ for a
  source distance of 10\,kpc, which could then in principle be
  explained by a symmetric supernova explosion or stellar diffusion.
  While a distance at the lower end of the allowed range would give a
  higher value for the current peculiar velocity, the peculiar
  velocity would still be sufficiently low at certain points in the
  Galactocentric orbit for an asymmetric kick not to be required.
  However, if the source is at the upper end of the possible distance
  range ($\gtrsim 12$\,kpc), the peculiar velocity is high enough
  throughout the orbit that an asymmetric kick becomes necessary.
  Without knowing the source distance, we cannot definitively
  determine its formation mechanism.  We also note that the momentum
  imparted by a kick of 23\,km\,s$^{-1}$ to such a $14M_{\odot}$ black
  hole \citep{Gre01} is similar to that gained by a neutron star
  receiving a kick of a few hundred km\,s$^{-1}$
  \citep[e.g.][]{Lyn94,Han97}, so owing to its large mass, even such a
  low peculiar velocity for GRS\,1915+105 would not necessarily rule
  out a natal kick.

For V404 Cyg, the derived components of the system velocity in
  the Galactic plane, $U$ and $V$, are much larger than $W$, the
  velocity out of the plane (Table \ref{tab:uvw}).  While this is to
  be expected since the Galactic rotation of 236\,km\,s$^{-1}$ forms a
  component of both $U$ and $V$, but not $W$, accounting for the
  Galactic rotation (giving the $U-<U>$ and $V-<V>$ terms in Table
  \ref{tab:uvw}) does not remove this discrepancy between the
  components of the peculiar velocity in and out of the plane.  A
  similar discrepancy is observed for the other four black hole
  systems with measured three-dimensional space velocities
  \citep{Mir01,Mir02,Mir03,Dha07}, even accounting for the ranges of
  values allowed by the uncertainties in the system parameters.
  However, as shown in Fig.~\ref{fig:velocities}, the velocity
  components change with time, so we reconstructed the Galactocentric
  orbits of all five sources.  This showed that the $W$ component of
  velocity can be significantly greater than the peculiar velocity in
  the plane for XTE J\,1118+480, and can be of a similar magnitude to
  the component in the plane at certain points in the orbits of Cygnus
  X-1 and GRS\,1915+105.  With only five systems, we are dealing with
  small number statistics, but these results are consistent with there
  being no preferred orientation of the peculiar velocity relative to
  the Galactic plane.  Indeed, considering the known population of
  black hole X-ray binaries, including those with no measured space
  velocity, the distribution in $z$ \citep{Jon04} demonstrates that a
  number of systems must have a significant $W$ velocity at certain
  points in their orbits, in order to reach the distances of several
  hundred parsecs above or below the plane at which they are currently
  observed.  Thus there is no observational evidence to suggest that
  the natal kick distribution should not be isotropic.

Thus the two systems with the lowest black hole masses, XTE
J\,1118+480 and GRO J\,1655-40 \citep[both of order
  6--7\,$M_{\odot}$][]{Oro03}, are found to have required asymmetric
kicks during the formation of the black hole.  For the higher-mass
systems, the situation is less clear-cut.  Cygnus X-1, V404 Cyg and
GRS\,1915+105 all have masses of $>10M_{\odot}$.  While Cygnus
  X-1 appears to have formed by direct collapse, the peculiar
  velocity of V404 Cyg implies that a supernova must have occurred,
  and an asymmetric kick seems likely to have been required in this
  system, contrary to assertions that black holes of $10M_{\odot}$
  form by direct collapse \citep[e.g.][]{Mir08}.  For GRS\,1915+105, we
  cannot definitively determine its formation mechanism without an
  accurate distance to the source.  Regardless, with only five systems,
no clear trends with black hole mass can be identified.  In order to
better constrain the formation mechanisms of stellar-mass black holes,
the space velocities of more such systems with a range of black hole
masses need to be measured, to constrain the frequency of occurrence
of natal kicks, and hence supernova explosions.

\section{Conclusions}
We have measured the proper motion of V404 Cyg using 20 years' worth
of VLA and VLBI radio observations.  Together with the radial velocity
and constraints on the distance of the system, this translates to a
peculiar motion of $64.1\pm3.7^{+37.8}_{-16.6}$\,km\,s$^{-1}$.  Given
the measured proper motion, the black hole cannot have been formed via
direct collapse.  A supernova is required to achieve the observed
peculiar velocity, with either a large amount of mass
($\sim11M_{\odot}$) being lost in the explosion, or, more probably,
the system being subject to an asymmetric kick.  In the case of a pure
Blaauw kick, $\sim 1M_{\odot}$ must have been transferred from the
donor to the black hole since the onset of mass transfer, implying an
initial black hole mass of $\sim 9M_{\odot}$ with a donor mass of
$\sim 2M_{\odot}$ prior to the onset of mass transfer.

\section*{Acknowledgments}
J.C.A.M.-J.  is a Jansky Fellow of the National Radio Astronomy
Observatory.  E.G. is supported through Chandra Postdoctoral
Fellowship grant number PF5-60037, issued by the Chandra X-Ray Center,
which is operated by the Smithsonian Astrophysical Observatory for
NASA under contract NAS8-03060.  The VLA and VLBA are facilities of
the National Radio Astronomy Observatory which is operated by
Associated Universities, Inc., under cooperative agreement with the
National Science Foundation.  The European VLBI Network is a joint
facility of European, Chinese, South African and other radio astronomy
institutes funded by their national research councils.  ParselTongue
was developed in the context of the ALBUS project, which has benefited
from research funding from the European Community's sixth Framework
Programme under RadioNet R113CT 2003 5058187.  This research has made
use of NASA's Astrophysics Data System.

\label{lastpage}
\bibliographystyle{mn2e}

\end{document}